\documentstyle[12pt]{article}

\newlength{\dinwidth}
\newlength{\dinmargin}
\def\lapproxeq{\lower .7ex\hbox{$\;\stackrel{\textstyle
<}{\sim}\;$}}
\def\gapproxeq{\lower .7ex\hbox{$\;\stackrel{\textstyle
>}{\sim}\;$}}
\def\funp{{I\!\!P}}
\def\be{\begin{equation}}
\def\ee{\end{equation}}
\def\bea{\begin{eqnarray}}
\def\eea{\end{eqnarray}}

\makeatletter
\def\fmslash{\@ifnextchar[{\fmsl@sh}{\fmsl@sh[0mu]}}
\def\fmsl@sh[#1]#2{%
\mathchoice
{\@fmsl@sh\displaystyle{#1}{#2}}%
{\@fmsl@sh\textstyle{#1}{#2}}%
{\@fmsl@sh\scriptstyle{#1}{#2}}%
{\@fmsl@sh\scriptscriptstyle{#1}{#2}}}
\def\@fmsl@sh#1#2#3{\m@th\ooalign{$\hfil#1\mkern#2/\hfil$\crcr$#1
#3$}}
\makeatother

\begin{document}
\titlepage
\begin{flushright}
DTP/97/12  \\
January 1997 \\
(revised March 1997)\\
\end{flushright}

\begin{center}
\vspace*{2cm}
{\Large \bf The rapidity gap Higgs signal at LHC} \\
\vspace*{1cm}
V.A.\ Khoze$^{a,b}$, A.D.\ Martin$^a$ and M.G.\ Ryskin$^{a,b}$
\end{center}

\vspace*{0.5cm}
\begin{tabbing}
$^1$xxxx \= \kill
\indent $^a$ \> Department of Physics, University of Durham,
Durham, DH1 3LE, UK. \\

\indent $^b$ \> Institute for Nuclear Physics, St. Petersburg,
SU-188 350 Gatchina, Russia.
\end{tabbing}

\vspace*{2cm}

\begin{abstract}
We compare the $WW$ and pomeron-pomeron fusion mechanisms for the
double-\linebreak diffractive production of a Higgs boson.  We
determine the
suppression of the \lq rapidity gap' pomeron-pomeron fusion
events due to QCD radiative effects.  In particular we use
leading $\log$ techniques to estimate the cross sections for both
exclusive and inclusive double-diffractive Higgs production at
LHC energies.  The same approach can be applied to the
double-diffractive central production of large $E_T$ dijets. 
These two processes provide one of the most justified
applications of
various aspects of leading logarithm QCD techniques.
\end{abstract}

\newpage
The biggest challenge facing the experiments at the forthcoming
very high energy proton-proton collider (LHC) is the search for
possible Higgs bosons.  The present estimates based on the
Standard Model and its minimal supersymmetric extension favour
the existence of an intermediate mass Higgs boson $(M_H
\lapproxeq 2 M_W)$ \cite{CAR}.  In this case the best signals,
which are based on the decay modes $H \rightarrow b\overline{b}$
or $\gamma\gamma$, will be extremely difficult to isolate from
the background.

One novel possibility to reduce the background is to study the
central production of the Higgs in events with a large rapidity
gap on either side.  Such rapidity gaps appear automatically if
the Higgs is produced via $WW$ boson fusion, $pp \rightarrow WW
\rightarrow H$ \cite{DKT,DKS}; recent developments are given in
\cite{BJ,BPZ,LE}.  There have also been several discussions about
the possible advantage of using a similar rapidity gap signal in
which the $W$ boson is replaced by the pomeron, $\funp$, see for
example \cite{BJ,SNS,BL,CH,MS,LM}.

The motivation of these $\funp\funp \rightarrow H$ studies starts
with the observation that at LHC energies $gg$ fusion is the
dominant mechanism with a Higgs production cross section up to a
factor of 10 larger than that of $WW$ fusion.  However, although
$WW$ and $gg$ fusion mechanisms appear to have a similar
structure, in $gg$ fusion the colour flow induces many
secondaries which completely fill the original (partonic level)
rapidity gaps.  For this reason $\funp\funp \rightarrow H$
mechanism has been proposed \cite{SNS,BL} instead of $gg
\rightarrow H$.  The idea is based on the hope that on the one
hand $\sigma (\funp\funp \rightarrow H)$ will be at least of the
order of a few percent\footnote{For each reaction the
short-distance process is $gg \rightarrow H$, where we regard $g$
in the case of $\funp\funp$ fusion as a gluonic constituent of
the pomeron.} of $\sigma (gg \rightarrow H)$, while on the other
hand the colour flow is screened in $\funp\funp$ fusion leading
to rapidity gaps.

In order to make estimates of $\sigma (\funp\funp \rightarrow H)$
it is necessary to invoke a model for the pomeron.  One
possibility is the non-perturbative approach of refs.\
\cite{SNS,BL}, which we call examples of the
\lq\lq soft" pomeron.  Another possibility is to consider the
perturbative QCD so-called \lq\lq hard" pomeron, see for example
\cite{MS,LM}.  The literature shows a wide range of predictions,
which may be expressed in terms of two extreme estimates.  The
\lq\lq soft" pomeron-like models give the upper extreme with
\be
\sigma_{\rm max} (\funp\funp \rightarrow H) \; \sim \; \sigma (gg
\rightarrow H) \: (\sigma_{\rm el}/\sigma_{\rm tot})^2
\label{eq:a1}
\ee
where the \lq\lq suppression" factor containing the elastic and
total $pp$ cross sections is the probability of having two
rapidity gaps, one either side of the Higgs.  The low extreme,
based on the \lq\lq hard" pomeron \cite{MS,LM}, is
\be
\sigma_{\rm min} (\funp\funp \rightarrow H) \; \sim \; \sigma (gg
\rightarrow H) \: (M_H^2 \sigma_{\rm tot})^{-2}
\label{eq:a2}
\ee
where now the \lq\lq suppression" factor is the probability to
have a point-like two-gluon configuration (with $\lambda \sim
1/M_H)$ in each pomeron so that they have sufficient chance to
fuse into the Higgs.  These simple estimates of the
suppression factor range from $10^{-1}$ to $10^{-12}$.  Although
naive, these results are in fact quite representative of the
range of values that may be found in the literature.

Let us start from the ordinary $gg \rightarrow H$ fusion process.
A relevant Feynman diagram for \lq rapidity gap' production is
shown in
Fig.~1, where the additional $t$ channel gluon is needed to
screen the
colour.  (The reason for the dashed and dotted gluon lines will
be 
explained below.)  Within this two-gluon exchange picture of the
pomeron it is clear that the most
optimistic scenario is first to assume that the gluon, which
screens the colour, does not couple to the Higgs,
and second, to assume that it has small virtuality $Q_T^2$ to
enhance the probability of screening via a large value of
$\alpha_S$.  This idea was invoked in an attempt to describe the
diffractive events in small $x$ deep inelastic scattering
\cite{BE}.  The simplest and most extreme prediction is given in
ref.\ \cite{BH}.  It was assumed that the \lq screening' gluon is
so soft that there is no suppression, apart from a factor of
$1/N_c^2$ which is the probability of forming a colour singlet
$gg$ $t$-channel state.  The perturbative realisation of the soft
screening approach has been studied for Higgs production
\cite{CH} and for dijet production \cite{BC}.

An important question, which has not yet been addressed in the
literature, concerns the probability of relatively hard gluon
emission coming from distance scales $\lambda \gapproxeq 1/M_H$
shorter than the characteristic transverse size $(\sim 1/Q_T)$ of
the pomeron at which the colour flow is screened.  Such gluons
could fill up the rapidity gaps.  The goal of the present paper
is to estimate the suppression of the rapidity gap events due to
these effects.  We will show that the typical values of $Q_T$ of
the \lq screening' gluon are indeed much smaller than $M_H$, but
nevertheless are sufficiently large for perturbative QCD to be
applicable.

Of course, there is also a suppression of rapidity gap events due
to parton-parton rescattering and to the possibility of multiple
(or \lq pile-up') interactions at high luminosities
\cite{DKS,BJ,LE,GLM}.  For example, a rough estimate of the
former suppression is \cite{RR}
$$
\left [1 \: - \: 2 (\sigma_{\rm el} + \sigma_{SD})/\sigma_{\rm
tot} \right ]^2 \; \sim \; 1 \: - \: 10\%
$$
depending on the value of the cross section, $\sigma_{SD}$, for
single diffraction.  These suppressions are common effects for
any Higgs production model, including $\funp\funp$ and $WW$
fusion, as well as for the background processes.  Such effects
will not be discussed further.

We calculate the rate of both exclusive and inclusive Higgs
production.  In the exclusive process, $pp \rightarrow ppH$, only
the Higgs and the recoil protons occur in the final state.  Due
to the presence of the proton form factors, the Higgs is produced
with small transverse momentum $q_T$.  We find that the
production cross section is negligibly small.  On the other hand
in the inclusive process, $pp \rightarrow X + {\rm gap} + H +
{\rm gap} + X^\prime$, the initial protons are destroyed.  The
phase space available for Higgs production, and hence $q_T$, are
large.  The cross section is found to be comparable to
that for $WW$ fusion.

We will work in the double logarithmic approximation (DLA) and
even for the Born amplitude we will use the leading power of
all logarithms to simplify the calculations.  Due to the large
value of the Higgs mass $M_H$, this approach is rather well
justified.  Indeed double-diffractive Higgs production provides
probably one of the most justified applications of various
aspects of leading $\log$ techniques to date. \\

\newpage

\noindent {\bf Exclusive production}

We start with the calculation of the (double-diffractive)
exclusive process which, at the quark level, is shown in Fig.~1. 
The Born amplitude for the process (shown by the solid lines in
the figure) is of the form
\be
M (qq \rightarrow q H q) \; = \; \frac{2}{9} \: 2A \: \int \:
\frac{d^2 Q_T}{Q^2 k_1^2 k_2^2} \: 4 \alpha_S^2 (Q^2) \:
(\mbox{\boldmath $k$}_1 . \mbox{\boldmath $k$}_2),
\label{eq:a3}
\ee
where $\frac{2}{9}$ is the colour factor for this colour-singlet
exchange process, and the factor of 2 takes into account that
both of
the $t$ channel gluons can emit the Higgs boson.
In the Standard Model the $gg \rightarrow H$
vertex factor is, after convolution with the gluon polarisations,
given by $A (\mbox{\boldmath $k$}_1 . \mbox{\boldmath $k$}_2)$
with
\be
A^2 \; = \; \sqrt{2} \: G_F \: \alpha_S^2 (M_H^2) \: N^2/9 \pi^2
\label{eq:a4}
\ee
where $G_F$ is the Fermi coupling, and where $N \approx 1$
provided that we are away from the threshold $M_H = 2m_t$.  Note
that in the forward direction [where $t_i = (q_i - q_i^\prime)^2
\rightarrow 0$ for $i = 1,2$ and $k_{1T} = k_{2T} = Q_T$] the
integral over the gluon loop reduces to $\int d^2 Q_T/Q^4$. 
Hence, as mentioned above, small values of $Q_T$ of the screening
gluon are favoured.

In order to make the (Born) calculation more realistic we first
have to include the ladder \lq evolution' gluons (shown by the
dashed lines in Fig.~1) and to consider the process $pp
\rightarrow pHp$ at the proton, rather than the quark, level. 
This is achieved by the replacements \cite{R,RRML}
\be
\frac{4 \alpha_S (Q^2)}{3 \pi} \: \rightarrow \: f (x, Q^2) \; =
\; \frac{\partial (xg (x, Q^2))}{\partial \ln Q^2}
\label{eq:a5}
\ee
where $x = x_1$ or $x_2$ for the upper or lower ladders in Fig.~1
respectively, and where $f (x, Q^2)$ is the unintegrated gluon
density\footnote{Strictly speaking even at zero transverse
momentum, $q_{1T} - q_{1T}^\prime = 0$, we do not obtain the
exact gluon structure function, as a non-zero component of
longitudinal momentum is transferred through the two-gluon
ladder.  However, in the region of interest, $x \sim 0.01$, the
value of $|t_{\rm min}| = m_p^2 x^2$ is so small that we may
safely put $t = 0$ and identify the ladder coupling to the proton
with the unintegrated gluon distribution $f (x, Q^2)$
\cite{RRML}.} 
of the proton.

The second correction to the Born formula, (\ref{eq:a3}), is the
inclusion of the Sudakov form factor $F_S$, which is shown
schematically
by the dotted curved line in Fig.~1.  $F_S$ is the probability
{\it not} to emit bremsstrahlung gluons (one of which is shown by
$p_T$) in the interval $Q_T \lapproxeq p_T \lapproxeq M_H / 2$. 
Clearly
the upper bound of the interval is $p_T \simeq M_H / 2$.
The lower bound, $Q_T \lapproxeq p_T$, occurs because there is
destructive
interference of the amplitude in which the bremsstrahlung gluon
is
emitted from a \lq hard' gluon $k_i$ with that in which it is
emitted from the soft \lq screening'
gluon $Q$.  That is there is no emission when $\lambda \simeq
1/p_T$ is larger than the separation, $\Delta \rho \sim 1/Q_T$,
of the two $t$-channel gluons in the transverse plane since then
they act as a single coherent colour-singlet system.
The Sudakov form factor for the above interval of $p_T$ is  
\be
F_S \; = \; \exp \biggl ( - S (Q_T^2, M_H^2) \biggr )
\label{eq:a6}
\ee
where $S$ is the mean multiplicity of bremsstrahlung gluons
\be
S (Q_T^2, M_H^2) \; = \; \int_{Q_T^2}^{M_H^2/4} \: \frac{C_A
\alpha_S (p_T^2)}{\pi} \: \frac{dE}{E} \: \frac{dp_T^2}{p_T^2} \;
= \; \frac{3 \alpha_S}{4 \pi} \: \ln^2 \left ( \frac{M_H^2}{4
Q_T^2} \right ).
\label{eq:a7}
\ee
Here $E$ and $p_T$ are the energy and transverse momentum of an
emitted
gluon in the Higgs rest frame.  The last equality assumes a fixed
coupling $\alpha_S$, and is shown only for illustration.  

Inserting corrections (\ref{eq:a5}) and (\ref{eq:a6}) into the
Born amplitude (\ref{eq:a3}) gives
\be
M (pp \rightarrow pHp) \; = \; A \pi^3 \: \int \:
\frac{dQ^2}{Q^4} \: e^{-S (Q_T^2, M_H^2)} \: f (x_1, Q_T^2) \: f
(x_2, Q_T^2)
\label{eq:a8}
\ee
in the leading $\log$ approximation.  The integral has a saddle
point given by
\be
\ln (M_H^2/4 Q^2) \; = \; (2 \pi/N_c \alpha_S (Q^2)) \: (1 - 2
\gamma)
\label{eq:b8}
\ee
where $\gamma$ is the anomalous dimension of the gluon, $g (x,
Q^2) \propto (Q^2)^\gamma$.  Suppose that we were to assume a
constant $\gamma = 0.15$.  Then for $M_H = 100$ (200) GeV the
saddle point would occur at $Q^2 = 7.3$ (14) GeV$^2$, well into
the perturbative region, and the Sudakov suppression of the cross
section would be $(F_S)^2 = 0.04$ (0.025).  However, a more
realistic evaluation using, say, the MRS(R2) set of partons
\cite{MRS} shows that the integrand reaches its maximum at $Q^2
\sim 2$ GeV$^2$, where the suppression is $(F_S)^2 = 0.003$
(0.0004).  If $\gamma$ were frozen in the region $Q^2 \leq 4$
GeV$^2$ then the cross section would be decreased by a further
factor of 2 --- a factor which is typical of the uncertainty.

Table 1 shows the values of the exclusive cross section,
\be
\frac{d \sigma}{dy} \: (pp \rightarrow p + H + p) \; = \;
\frac{|M|^2}{16^2 \pi^3 b^2},
\label{eq:c8}
\ee
calculated from (\ref{eq:a8}).  We have integrated over the
$dt_i$ assuming form factors $\exp (-bt_i/2)$ at the
proton-pomeron vertices, with $b = 5.5$ GeV$^{-2}$.  We find that
the cross section is more than a factor of $10^5$ smaller than
the inclusive
$pp \rightarrow gg \rightarrow H$ cross section, without rapidity
gaps; and even a factor 10 less than the $\gamma\gamma
\rightarrow H$ cross section \cite{PP}.  Exclusive
double-diffractive Higgs production is thus only of academic
interest. \\

\noindent {\bf Inclusive production}

We find that the cross section for inclusive double-diffractive
Higgs production is much larger.  Here the initial protons may be
destroyed and the transverse momentum of the Higgs is no longer
limited by the proton form factor, and so the Sudakov suppression
is weaker.  The process is shown in Fig.~2 in the form of the
amplitude multiplied by its complex conjugate.  The partonic
quasielastic subprocess is $ab \rightarrow a^\prime + {\rm gap} +
H + {\rm gap} + b^\prime$.  If the partons $a,b$ are quarks then
the Born amplitude for the subprocess is given by (\ref{eq:a3}). 
However, the form factor suppressions are more complicated than
for the exclusive process.  As the momenta transferred, $t_i = (Q
- k_i)^2$, are large we can no longer express the upper and lower
\lq blocks' in terms of the gluon structure function, but instead
they are given by BFKL non-forward amplitudes.

We begin with the expression for the Born cross section for the
subprocess $gg \rightarrow g + H + g$
\be
\frac{d\sigma}{dy} \; = \; A^2 \: \alpha_S^4 \: \frac{81}{2^{8}
\pi} \: {\cal I}
\label{eq:a9}
\ee
with
\be
{\cal I} \; = \; \frac{1}{\pi^2} \: \int \: \frac{dQ^2}{Q^2} \:
\frac{dQ^{\prime 2}}{Q^{\prime 2}} \: \frac{d^2 k_{1T}}{k_1^2
k_1^{\prime 2}} \: \frac{d^2 k_{2T}}{k_2^2 k_2^{\prime 2}} \:
(\mbox{\boldmath $k$}_{1T} . \mbox{\boldmath $k$}_{2T})
(\mbox{\boldmath $k$}_{1T}^\prime . \mbox{\boldmath
$k$}_{2T}^\prime ),
\label{eq:a10}
\ee
where the six propagators of Fig.~2 are evident.  As before the
leading $\log$ contribution comes from the region where the
screening gluons are comparatively soft.  That is $Q_T \ll
k_{iT}$ and $Q_T^\prime \ll k_{iT}^\prime$, and so
\be
t_i \; = \; (Q - k_i)^2 \: \simeq \: - k_{iT}^2 \: \simeq \:
- k_{iT}^{\prime 2}
\label{eq:a11}
\ee
for $i = 1,2$.  After performing the azimuthal integrations, the
$gg \rightarrow H$ vertex factors become
\be
(\mbox{\boldmath $k$}_{1T} . \mbox{\boldmath
$k$}_{2T})(\mbox{\boldmath $k$}_{1T}^\prime . \mbox{\boldmath
$k$}_{2T}^\prime) \: \rightarrow \: \textstyle{\frac{1}{2}} \:
k_{1T}^2 \: k_{2T}^2 \; \simeq \; \textstyle{\frac{1}{2}} \: t_1
t_2
\label{eq:a12}
\ee
and we see that (\ref{eq:a10}) indeed yields the maximum number
(four) of logarithms.

Again we must estimate the suppression due to gluon
bremsstrahlung filling up the rapidity gaps.  Now the mean number
of gluons emitted, with transverse momenta $Q_T < p_T < k_{iT}$,
in the rapidity interval $\Delta \eta_i$ is
\be
n_i \; = \; \frac{\alpha_S N_c}{\pi} \: \Delta \eta_i \: \ln
\left ( \frac{k_{iT}^2}{Q_T^2} \right ).
\label{eq:a13}
\ee
The amplitude for no emission in the gap $\Delta \eta_i$ is
therefore $\exp (-n_i/2)$.  In this way we see that the Born
integral (\ref{eq:a10}) is modified to
\be
{\cal I} \; = \; \frac{1}{2} \: \int \: \frac{dQ^2}{Q^2} \:
\frac{dQ^{\prime 2}}{Q^{\prime 2}} \: \frac{dt_1}{t_1} \:
\frac{dt_2}{t_2} \: \exp \left (- \frac{n_1}{2} -
\frac{n_1^\prime}{2} - \frac{n_2}{2} - \frac{n_2^\prime}{2} -
S - S^\prime \right )
\label{eq:a14}
\ee
where the exponential factor represents the total form factor
suppression in order to maintain the rapidity gaps $\Delta
\eta_1$ and $\Delta \eta_2$ in Fig.~2.  The Sudakov form factor,
$\exp (- S (k_T^2, M_H^2))$, arises from the insistence that
there is no gluon emission in the interval $k_T < p_T <
\frac{1}{2} M_H$, see (\ref{eq:a6}) and (\ref{eq:a7}).

The justification of the non-Sudakov form factors, $\exp
(-n_i/2)$ is a little subtle.  First we notice from
(\ref{eq:a13}) that due to the asymmetric configuration of the
$t$-channel gluons, $Q_T \ll k_{iT}$, we have, besides $\Delta
\eta_i$, a second logarithm, $\ln (k_{iT}^2/Q_T^2)$, in the BFKL
evolution.  These double $\log$s are resummed\footnote{The
resummation corresponds to the Reggeization of the $t$-channel
gluons.} to give the BFKL non-forward amplitude $\exp (-n_i/2)
\Phi (Y_i)$, where the remaining factor $\Phi (Y_i)$ accounts for
the usual longitudinal BFKL logarithms\footnote{Here $\Delta
\eta_i$ (or $Y_i$) plays the role of $\ln (1/x)$ in the BFKL
evolution.},
\be
Y_i \; \equiv \; (\alpha_S N_c/2 \pi) \Delta \eta_i.
\label{eq:a15}
\ee
In the region of interest at LHC energies, $Y_i \lapproxeq 0.3$,
it is sufficient to include only the ${\cal O} (Y_i)$ term, which
gives $\Phi \simeq 1 + Y_i Q_T^2/k_{iT}^2 \simeq 1.1 \pm 0.1$
\cite{FR}.  At our level of accuracy we may neglect the
enhancement due to $\Phi$, and hence we obtain (\ref{eq:a14}),
which is valid in the double $\log$ approximation.

To evaluate ${\cal I}$ of (\ref{eq:a14}) we first perform the
$Q^2$ and $Q^{\prime 2}$ integrations and obtain $(Y_1 + Y_2)^{-
2}$.  Then we integrate over $\ln (t_1/t_2)$ which gives
$\frac{1}{2} (1/Y_1 + 1/Y_2)$ where, at large $\Delta \eta_i$, we
neglect the $t_i$ dependence of $S_i$.  Thus (\ref{eq:a14})
becomes
\be
{\cal I} \; = \; \frac{1}{4 Y_1 Y_2 (Y_1 + Y_2)} \:
\int^{\frac{1}{4} M_H^2} \: \frac{dt}{t} \: \exp (-2 S (t,
M_H^2)).
\label{eq:a16}
\ee
For fixed $\alpha_S$ the final $(dt)$ integration gives $\pi (2
N_c \alpha_S)^{- \frac{1}{2}}$ in the DLA.  However, to predict
the cross section for inclusive production at the LHC we must
convolute the parton-parton cross sections with the parton
densities $a (x_i, t)$ of the proton, with $a = g$ or $q$, and
evaluate the $dt$ integral numerically.  There is a subtlety when
we come to include these parton luminosity factors
$$
\int_{x_{\rm min}}^1 \: dx_a \: a (x_a, k_{1T}^2) \: \ldots ,
$$
with $a = g, q$.  At first sight we might expect $x_{\rm min} =
x_H \equiv M_H/\sqrt{s}$ for central Higgs production.  However,
at large $k_{iT}$ the rapidities of the $a^\prime, b^\prime$ jets
are small in the Higgs rest frame; $\eta_{a^\prime} = \ln
(x_{a^\prime} \sqrt{s}/k_{1T})$.  Thus in order to maintain the
rapidity gaps $(\eta_{a^\prime} > \Delta \eta_1)$, we must take
\be
x_{\rm min} \; = \; x_H (1 + k_{1T} \exp (\Delta \eta_1)/M_H).
\label{eq:a17}
\ee

The results for $\sigma_{\rm in} (\funp\funp)$ shown in table 1
are the sum over all types of initial partons, and correspond to
$\Delta \eta_1 = \Delta \eta_2 = \Delta \eta$ where $\Delta
\eta$, the parton level rapidity gap, is taken to be either
$\Delta \eta = 2$ or $\Delta \eta = 3$.  From
(\ref{eq:a16}) we see that the rapidity gap cross section
decreases as $1/Y^3$, that is as $(1/\Delta \eta)^3$, if $\Delta
\eta_1 = \Delta \eta_2$.  As expected, the suppression decreases
with increasing $\alpha_S$ (like $\alpha_S^{-3.5}$ in the DLA). 
For comparison we give the estimates for the $WW \rightarrow H$
cross section for the same rapidity gap configuration.  From
table 1 we see that the $\funp\funp$ fusion Higgs signal is
comparable to that of $WW$ fusion for $M_H \simeq {\rm 100 GeV}$,
but
is of decreasing importance as the value of $M_H$ (or $\Delta
\eta$)
increases.  However
NLO corrections (which are not included in table 1) may increase
the value of $\sigma_{\rm in} (\funp\funp \rightarrow H)$ by a
factor\footnote{The $K$ factor enhancement (analogous to that in
Drell-Yan production) is expected to be 1.6--2 \cite{KLS}, and
there could be a factor of up to 2 from the single $\log$ BFKL
enhancement term $\Phi^4$.} of 2--4. 

It is interesting to note that, due to the strong ordering of
$k_T$ in the leading $\log$ approximation, almost all the
momentum transfer $k_{iT}$ is balanced by the $k_{jT}$ of the
parton which borders the rapidity gap.  Thus, in principle, the
$\funp\funp$ and $WW$ signals could be distinguished by the
transverse momentum $k_{jT}$ of the jets which border the
rapidity gaps\footnote{For $WW$ fusion one half of the cross
section comes from events with $k_{jT} \lapproxeq M_W$, while for
$\funp\funp$ fusion one half comes from $k_T < 13$ (25) GeV for
$M_H = 100$ (300) GeV if $\Delta \eta = 2$.  Indeed we could
reduce $WW$ fusion in comparison with $\funp\funp$ fusion by
about a factor $(k_{jT}^2/M_W^2)^2$ by selecting events with both
\lq border' jets satisfying the cut-off $k_{jT}^2 < k_0^2$, with
$k_0^2$ taken to be much less than $M_W^2$.  For example, for
$M_H = 100$ GeV, $\Delta \eta = 2$ if we take $k_0 = 20$ GeV
then we find $d \sigma_{\rm in} (\funp\funp \rightarrow H)/dy =
200$ fb as
compared to $d\sigma_{\rm in} (WW \rightarrow H)/dy = 4$ fb.}.  
Also note that we have implicitly assumed that there is no 
interference between the $\funp \funp \rightarrow H$ and $WW
\rightarrow
H$ amplitudes.  This is a good approximation since the two
amplitudes
(i) are essentially out of phase, (ii) produce Higgs with
different
$q_T$ distributions, and (iii) involve different partons (namely
$a = g$ for $\funp \funp$ fusion and $a = q$ for $WW$ fusion.) \\

\noindent {\bf Discussion}

Recall that our DLA approach to $\funp\funp \rightarrow H$ is
only justified for the asymmetric configuration of the $t$
channel gluons, $Q_T^2 \ll k_{iT}^2$.  We must check that this is
in fact the case.  We have seen above that typically $k_{iT} \sim
20$ GeV at LHC energies.  Now, taking $\alpha_S = 0.2$, we have
$Y \simeq 0.1 \Delta \eta \simeq 0.25$.  Thus, using
(\ref{eq:a13}), we find $\ln (k_{iT}^2/Q_T^2) \simeq 1/2Y \simeq
2$.  So indeed $Q_T^2 < 0.15$ $k_{iT}^2$.  Since $k_{iT}$ is
rather large, the suppression due to the Sudakov form factor is
not so strong for inclusive production, $(F_S)^2 \simeq 0.5$.  We
conclude that for relatively small $Y$, say $Y < 0.3$, the
approach is self-consistent and we may use the DLA expressions,
$\exp (- Y_i \ln (k_{iT}^2/Q_T^2))$, for the BFKL non-forward
amplitudes, see (\ref{eq:a13}) and (\ref{eq:a14}).  Moreover we
have seen that the suppression has a clear physical
interpretation.

At large $Y$, say $\Delta \eta > 5$, the situation is different. 
As $\Delta \eta$ increases we enter the symmetric BFKL gluon
configuration, $Q_T^2 \sim k_{iT}^2$.  We no longer have double
$\log$s (and moreover we lose three $\log$s from the $Q^2,
Q^{\prime 2}$ and $d (t_1 - t_2)$ integrations in ${\cal I}$ of
(\ref{eq:a14})).  Instead, at large $Y$ and $t \neq 0$, we have
the familiar exponential growth of the BFKL amplitude arising
from the resummation of the (single) longitudinal
$\log$s\footnote{For example, in a recent study \cite{FR} of
$J/\psi$ electroproduction at large $t$ (but in the symmetric
gluon configuration $Q_T^2 \sim k_{iT}^2$) it was found for $Y =
0.35$ that the factor $\Phi (Y)$ enhanced the cross section by
about a factor of 5; see also \cite{CH}.}.  We obtain
\be
\Phi (Y) \; \sim \; \exp (\lambda \Delta \eta)/(\Delta
\eta)^{\frac{3}{2}}
\label{eq:a18}
\ee
where $\lambda$ is the BFKL intercept.  Due to the $\Delta \eta$
term in the denominator, the growth only starts at $\Delta \eta
\sim \frac{3}{2} \lambda^{-1} \gapproxeq 5$ (if we take $\lambda
\sim 0.3$ from the rise of $F_2$ observed at HERA with decreasing
$x$).  This rapidity gap configuration is beyond the LHC energy
range and is not discussed further here, although it could become
important at very high energies.

Our conclusion is that the interesting proposal \lq\lq that the
Higgs signal could be improved by studying production in the
double-diffractive configuration" does not look so optimistic as
it first seemed.  Exclusive production is negligibly small and
even the inclusive cross section is of the same order as the
cross
section for the more familiar $WW \rightarrow H$ process.  The
problem is that QCD radiation has a large probability to fill the
parton-level rapidity gaps.

Finally we note that our approach may be used to estimate the
cross section for the central production of a pair of high $E_T$
jets with a rapidity gap either side of the pair.  We simply need
to replace the $gg \rightarrow H$ cross section by that for $gg
\rightarrow$ dijet.  Since the latter cross section is much
larger, and since we have an extra parameter $E_T$, such dijet
production (even at Fermilab energies) offers an excellent
opportunity to study QCD (double and single) leading $\log$
techniques.  Moreover, estimates of this dijet production will be
important to determine the level of the background to the $H
\rightarrow b\overline{b}$ signal. \\

\noindent {\bf Acknowledgements}

We thank John Ellis and Peter Landshoff for useful discussions. 
VAK thanks PPARC for support and MGR thanks the Royal Society,
INTAS (95-311) and the Russian Fund of Fundamental Research (96
02 17994) for support.

\begin{table}[htb]
\caption{The cross sections $d\sigma/dy|_0$ (in fb) for the
double-diffractive central production of a Higgs boson in $pp$
collisions at ${\protect \sqrt{s}} = 14$ TeV via $\funp\funp$ or
$WW$ fusion.  Here $\funp$ denotes the \lq hard' QCD pomeron, and
$\sigma_{\rm ex, in}$ refer to exclusive, inclusive production
respectively.  The MRS(R2) set of partons {\protect \cite{MRS}}
is used.  We take running $\alpha_S$ in the evaluation of the
Sudakov form factors, with $\alpha_S (M_Z^2) = 0.118$, but fixed
$\alpha_S = 0.2$ in the BFKL amplitude.}
\begin{center}
\begin{tabular}{|c|c|c|c|} \hline
$M_H$ (GeV) & $\sigma_{\rm ex} (\funp\funp)$ & $\sigma_{\rm in}
(\funp\funp)$ & $\sigma_{\rm in} (WW)$ \\
& & $\Delta \eta = 2$ (3) & $\Delta \eta = 2$ (3) \\ \hline
100 & $18 \times 10^{-2}$ & 300 (33) & 220 (60) \\
200 & $5 \times 10^{-3}$ & 85 (9) & 180 (50) \\
300 & $4.4 \times 10^{-4}$ & 36 (4) & 140 (40) \\ \hline
\end{tabular}
\end{center}
\end{table}

\medskip
\noindent {\large \bf Figure Captions}
\begin{itemize}
\item[Fig.~1] The Born amplitude for the exclusive
double-diffractive production of a Higgs boson of transverse
momentum $q_T$, shown together with the QCD radiative corrections
arising from \lq evolution' gluons (dashed lines) and the Sudakov
form factor (curved dotted lines).  The \lq soft screening' gluon
has four-momentum $Q$.

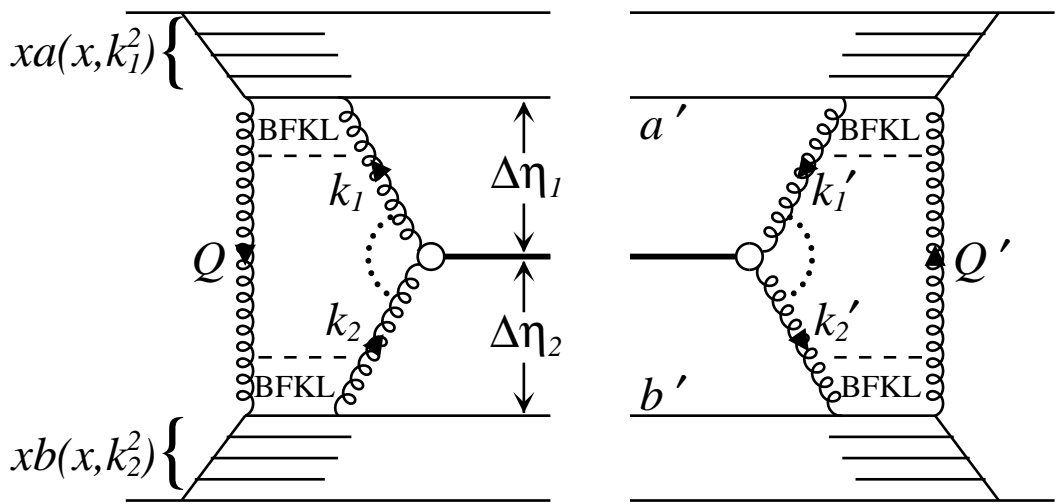
\item[Fig.~2] The amplitude multiplied by its complex conjugate
for the inclusive central production of a Higgs boson with
rapidity gaps $\Delta \eta_1$ and $\Delta \eta_2$ on either side.
The suppression due to QCD radiative effects comes from the
double $\log$ resummations $\exp (-n_i/2)$ in the BFKL
non-forward
amplitudes and from the Sudakov form factors $\exp (-S)$ shown by
the dotted curves; see eq.\ (\ref{eq:a14}).

\end{itemize}

\end{document}